# Wage Rigidity, Exchange Rate Regimes, and Inflation Persistence in Transition Economies: A Cohort-Based Institutional Approach

**Stefan Tanevski**
University American College Skopje
stefan.tanevski@uacs.edu.mk

**Marjan Petreski**
University American College Skopje
marjan.petreski@uacs.edu.mk

## Abstract

This paper investigates how institutional rigidities shape inflation persistence in transition economies, focusing on labor market institutions and exchange rate regimes. Using a large panel of transition countries over the period 2013–2024, the analysis combines newly constructed indices of wage rigidity and labor protection—derived from AI-assisted coding of legal texts—with de facto measures of exchange rate regime rigidity and standard macroeconomic controls. The empirical strategy adopts a dynamic panel framework in which inflation persistence is conditioned on institutional characteristics through interaction terms, estimated using GMM techniques. Identification follows a cohort-based approach, comparing inflation dynamics across countries with different institutional configurations. To address potential measurement and classification uncertainty in institutional variables, the analysis incorporates a simulation-based sensitivity framework. The results show that inflation persistence varies systematically across institutional settings. Both wage rigidity and exchange rate regime rigidity tend to dampen inflation persistence, indicating that institutional constraints can weaken the transmission of past inflation into current price dynamics. This effect is particularly strong and robust for exchange rate regimes, while the effect of wage rigidity is more sensitive to measurement assumptions. Findings highlight the importance of institutional structures in shaping inflation processes and suggest that nominal rigidities may play a stabilizing role in certain macroeconomic environments.

## 1. Introduction

Nominal rigidities are conventionally treated as amplifiers of inflation persistence. In the standard New Keynesian framework, sluggish wage and price adjustment slows the response of real marginal costs to shocks, sustains backward-looking inflation dynamics, and prolongs the return of inflation to target (Taylor, 1980; Calvo, 1983; Christiano et al., 2005). The implication is straightforward: stronger rigidities in wage-setting and more constrained exchange-rate adjustment should increase, rather than reduce, inflation persistence.

Whether that prediction carries over to transition economies is, however, not obvious, and the reason lies in the institutional history of wage formation in these economies. Wage-setting under socialism was administratively structured through occupational and sectoral grids tied more closely to political and distributive objectives than to productivity or market conditions, and wage expectations were shaped by those administrative arrangements rather than by inflation-responsive bargaining (Vodopivec, 1991; Milanovic, 1998). Collective bargaining institutions, where they existed, were only weakly independent and did not operate through the adversarial, inflation-responsive logic characteristic of advanced market economies (Boeri and Terrell, 2002; Svejnar, 2002). Labour-market adjustment during transition was further shaped by fragmented bargaining structures, protected incumbency, and a gradual shift away from compressed wage systems rather than by an immediate move to market-clearing wage formation (Boeri and Terrell, 2002; Izyumov and Vahaly, 2009).

These features did not dissolve at the onset of transition; they persisted as institutional legacies conditioning expectations and adjustment behaviour well into the consolidation phase. The analytical implication is not simply that wages adjusted slowly. It is that measured wage rigidity in this setting need not reinforce backward-looking indexation; it may instead capture an inherited institutional architecture in which the transmission of past inflation into current wage claims was attenuated from the outset. Two further structural features compound this. Exchange-rate regimes across the region have frequently been used as explicit anti-inflation commitment devices rather than passive nominal arrangements, and where credible, they discipline expectations by tying nominal stabilisation to an external anchor rather than to the history of domestic inflation (Calvo and Reinhart, 2002; Fischer, 2001; Calvo, Celasun and Kumhof, 2002). At the same time, these economies are small, highly open, and disproportionately exposed to imported energy and traded-goods prices, so that external cost shocks dominate the inflation process and leave limited room for the domestic wage channel to generate or sustain second-round effects. Under these conditions, wage-price feedback may be institutionally attenuated, expectations may be anchored by the exchange-rate regime, and inflation dynamics may be dominated by external-price shocks. The channels through which nominal rigidities are expected to amplify inflation may therefore be weaker, differently configured, or jointly constrained.

The question this paper pursues is whether the interaction between nominal rigidities and inflation persistence takes a different form in transition economies, where the institutional basis of wage formation, the credibility function of exchange-rate regimes, and the structure of the inflation process all differ materially from the advanced-economy settings. Using a dynamic panel GMM framework in which inflation persistence is conditioned on institutional characteristics, and exploiting cross-country heterogeneity in institutional configurations under a cohort-based identification strategy, the analysis finds that both wage and exchange-rate regime rigidity are associated with lower inflation persistence, not higher. The exchange-rate result is particularly strong and remains highly robust once classification uncertainty in the de facto regime assignment is articulated through simulation. The wage-rigidity effect is directionally similar but more sensitive to measurement assumptions, which is consistent with the fact that the index captures the institutional structure of wage formation rather than backward-looking indexation behaviour directly.

These findings run against the conventional prediction of the New Keynesian literature, but they are consistent with a coherent set of transition-economy mechanisms. Institutionally administered wage-setting may attenuate, rather than reinforce, the pass-through from past inflation to current wage claims. Exchange-rate arrangements that operate as nominal anchors may discipline expectations and weaken the backward-looking component of inflation. And where external price shocks dominate the inflation process, the domestic wage–price loop may be truncated before institutional rigidities have an opportunity to sustain it. In such an environment, the conditions under which nominal rigidities amplify persistence are jointly weakened.

The paper makes four contributions. First, it provides new cross-country evidence that institutional rigidities can dampen, rather than amplify, inflation persistence in transition economies, showing that the standard New Keynesian amplification result need not hold under the institutional and structural conditions that characterize this setting. Second, it constructs original, legally grounded indices of wage rigidity and labour protection, derived from AI-assisted coding of labour codes, collective-bargaining legislation, and minimum-wage provisions, thereby providing transparent and replicable measures of the de jure institutional architecture of wage-setting that are absent from existing cross-country sources for this set of economies. Third, it adopts a cohort-based identification strategy that exploits cross-country heterogeneity in institutional configurations rather than relying on within-country institutional variation, which is limited by design given the slow-moving nature of labour-market legislation and the discrete character of exchange-rate regime changes. Fourth, it develops a simulation-based sensitivity framework that treats coding uncertainty in the labour-market indices and classification uncertainty in exchange-rate regimes as conceptually distinct measurement problems, and propagates both through the full dynamic panel GMM procedure.

The policy implication is not that rigidities are generically stabilising. It is narrower. In economies where wage formation is rule-bound, exchange-rate regimes perform a credibility function, and inflation is driven primarily by external cost shocks, institutional constraints may dampen rather than intensify inflation persistence. The relevant object for stabilisation policy is therefore not any single rigidity in isolation, but the joint configuration of the internal and external adjustment margins through which inflation is propagated.

The remainder of the paper is organised as follows. Section 2 reviews the literature on inflation persistence, nominal rigidities, exchange-rate regimes, and inflation dynamics in transition economies, and positions the paper's contribution within that landscape. Section 3 presents few stylized facts. Section 4 describes the construction of the institutional indices, the exchange-rate classification, and the treatment of measurement uncertainty, as well sets out the empirical specification, the cohort-based identification strategy, and the simulation framework. Section 5 presents and interprets the baseline results, simulation evidence, and robustness checks. Section 6 concludes.

## 2. Literature Review

Within the conventional New Keynesian literature, nominal rigidities are typically modelled as amplifiers of inflation persistence. The benchmark mechanism runs through sluggish wage and price adjustment, which slows the response of real marginal costs to shocks and sustains backward-looking inflation dynamics in price-setting (Taylor, 1980; Calvo, 1983; Christiano et al., 2005). Implicit in this benchmark is a wage-setting environment that remains fundamentally market-responsive, so that rigidity preserves the pass-through from past inflation into current wage claims rather than suppressing it altogether.

The degree of inflation persistence varies systematically across countries, time periods, and policy regimes, a pattern that has generated a substantial empirical literature seeking to identify its structural determinants. Fuhrer (1997, 2010) and Galí and Gertler (1999) document that backward-looking or sluggishly adjusting inflation expectations are a primary source of persistence, while Clarida, Galí and Gertler (1999), Levin and Piger (2004), and Stock and Watson (2007) show that the credibility of the monetary-policy framework is a key conditioning variable: stronger and more systematic policy responses are associated with lower persistence, while accommodative or weak frameworks prolong inflationary episodes. This literature establishes that inflation persistence is not a fixed structural parameter but an endogenous outcome shaped by the interaction between nominal rigidities, expectations formation, and the institutional environment in which monetary policy operates.

A smaller but theoretically important literature indicates that the relationship between wage rigidity and inflation dynamics need not be monotone. Blanchard and Galí (2007) show, within a tractable New Keynesian framework, that real wage rigidities can alter marginal-cost dynamics in ways that do not necessarily intensify inflation persistence under all parameterizations. Their argument is not that wage rigidity is generically disinflationary, but that once real wages become less responsive to cyclical conditions, the response of marginal costs to aggregate-demand shocks may itself be attenuated. Under certain combinations of price and wage stickiness, this weakens rather than strengthens inflation propagation. Related contributions reach a similar conclusion in adjacent settings. Benigno and Ricci (2011) show that downward nominal wage rigidity alters the inflation–output trade-off as a function of trend inflation, implying that the persistence effects of wage rigidity are conditional on the macroeconomic environment rather than uniform across settings. Christoffel and Kuester (2008) demonstrate that the inflation consequences of wage stickiness depend critically on how bargaining enters the price-setting mechanism, again pointing to transmission structure rather than rigidity per se as the determinant of persistence. The common implication is that the sign and magnitude of the rigidity–persistence relationship are not theoretically settled; they depend on the institutional form that rigidity takes and the environment in which it operates.

This qualification is especially relevant in transition economies because the institutional basis of wage formation differs materially from the market-clearing benchmark assumed in much of the standard literature. In these settings, wage determination often operated through institutionally administered rather than purely competitive mechanisms, including public-sector pay grids, hierarchical bargaining arrangements, partial legal extension of collective agreements, and policy-driven incomes frameworks (Vodopivec, 1991; Svejnar, 1992; Boeri and Terrell, 2002). The analytical implication is that measured wage rigidity may capture a weaker mapping from past inflation to current wage claims rather than a stronger one. In that case, rigidity does not reinforce backward-looking wage indexation; it constrains it. The relevant mechanism is therefore not slower adjustment in an otherwise market-responsive wage-setting process, but a weaker pass-through from past inflation into current nominal wages.

Evidence from transition economies supports this interpretation. Svejnar's (1992) work on labour-market adjustment in transition economies emphasizes the role of wage controls, state-enterprise behaviour, and weakly marketized labour relations in shaping adjustment outcomes. Related evidence from Bulgaria indicates that non-competitive forces dominated wage determination under the planning regime and remained influential in early transition even as wage-setting became gradually more market-oriented (Jones and Simon, 2005). More broadly, labour-market adjustment during transition was conditioned by inherited institutional constraints and a gradual shift away from compressed wage structures rather than by an immediate move to market-clearing wage formation (Boeri and Terrell, 2002; Milanovic, 1998; Izyumov and Vahaly, 2009). These findings are relevant not because they establish a direct inflation result, but because they clarify the institutional content of wage rigidity in transition settings: rigidity here is not simply contract staggering in the New Keynesian sense, but part of a broader institutional architecture that may weaken backward-looking wage adjustment.

The literature on exchange-rate regimes offers a parallel qualification to the conventional view. While rigid exchange-rate arrangements are often criticized for foreclosing nominal exchange-rate adjustment, they have also been studied as commitment devices in economies with weak monetary credibility. Calvo and Reinhart (2002) document the phenomenon of "fear of floating," whereby countries that formally claim to float in fact intervene heavily to stabilize the exchange rate. Their interpretation is that exchange-rate stability often serves a credibility-building role in economies with histories of inflation or financial fragility. Fischer (2001) similarly concludes that, for small open economies with weak monetary institutions, hard pegs and tightly managed floats have often been associated with lower and more stable inflation, especially during disinflation phases.

The analytical relevance of this literature lies in the distinction between exchange-rate rigidity as a constraint on relative-price adjustment and exchange-rate rigidity as a nominal anchor. In the first interpretation, rigid regimes shift shock absorption onto domestic prices and therefore increase persistence. In the second, they discipline expectations by tying nominal stabilization to an external anchor rather than to the history of domestic inflation. Calvo, Celasun and Kumhof (2002) formalize this latter mechanism and show that exchange-rate anchors can break inflation inertia in small open economies with imperfect monetary credibility through the expectations channel. Under such conditions, the backward-looking component of inflation is weakened, and persistence may decline rather than rise.

A further qualification arises from the structure of inflation in transition economies themselves. Existing studies generally find that inflation in CESEE and CIS economies is strongly shaped by external price pressures, administered-price adjustments, and the credibility of the stabilization framework (Coricelli and Jazbec, 2004; Égert, 2007; Brada and Kutan, 2002). This implies that the domestic wage channel may carry less weight in inflation dynamics than in the advanced economies that underpin much of the standard New Keynesian literature. Put differently, the inflation process in transition economies is characterized by strong external-price pass-through and a comparatively weak domestic cost channel. These economies are small, highly open, and disproportionately exposed to imported energy, traded-goods prices, and imported intermediate inputs. Under such conditions, imported shocks dominate the dynamics of inflation, while wage-based second-round effects are structurally limited. For a wage–price spiral to generate substantial persistence, domestic wages must respond strongly to past inflation and domestic labour costs must account for a sufficiently large share of marginal-cost variation. Neither condition can be assumed in small open transition economies. External-price dominance therefore truncates the wage–price loop before labour-market institutions even enter the analysis. This structural feature complements the institutional argument rather than competing with it: institutionally constrained wage-setting weakens the backward-looking wage response, while external-price dominance reduces the aggregate importance of that response for inflation dynamics (Coorey, Mecagni and Offerdal, 1996; Boeri and Terrell, 2002)

Despite the relevance of these mechanisms, the literature remains fragmented in ways that leave the central question of this paper unanswered. Research on labour markets in transition economies has largely examined institutions through the lens of labour-market performance—focusing on wages, employment, unemployment, and reform episodes—rather than on their implications for inflation persistence (Boeri and Terrell, 2002; Brada and Slaveski, 2012; Bah and Brada, 2014; Izyumov and Vahaly, 2009; Petreski and Tanevski, 2024). Closest to the present paper are studies that relate labour-market institutions to labour-market outcomes rather than to inflation dynamics directly. Boeri and Terrell (2002), for example, explain divergent transition paths through institutional differences in wage floors, employment adjustment, and labour reallocation, while Bah and Brada (2014) review how labour-market flexibility, wages, privatization, and restructuring shaped labour-market performance across the region. In parallel, the literature on inflation in emerging and transition economies has focused mainly on exchange-rate pass-through, stabilisation regimes, and external price shocks, treating labour-market institutions as background conditions rather than as explicit conditioning variables in the persistence mechanism (Ghosh et al., 2014; World Bank, 2018).

The paper therefore examines whether the institutional and structural features of transition economies attenuate the propagation channels through which nominal rigidities are conventionally expected to amplify inflation persistence.

## 3. A Glimpse at Nominal Rigidities in Transition Economies

We build an intuition about the wage and exchange rate rigidities through few simple graphs. Figure 1 plots annual inflation against annual nominal wage growth across the sample of transition economies. The relationship is positive, but visibly loose and highly dispersed. At inflation rates in the 0–10 percent range, wage growth spans negative values, modest increases, and substantial accelerations. Even during higher-inflation episodes, wage responses vary considerably across observations. Nominal wages therefore do not appear to adjust one-for-one with inflation in any systematic way across the sample.

The descriptive implication is not that wage rigidity necessarily sustains inflation persistence, but that the domestic wage channel is heterogeneous and incomplete. Inflation shocks do not mechanically translate into proportional wage increases, and similar inflation outcomes are associated with markedly different wage responses. From a macroeconomic perspective, this suggests that wage adjustment is not a uniform transmission mechanism through which past inflation is carried into current labour costs. The strength of the wage–price link is therefore an empirical question rather than a maintained assumption.

Figure 1. Nominal Wage Growth and Annual Inflation Rate (2011-2024)

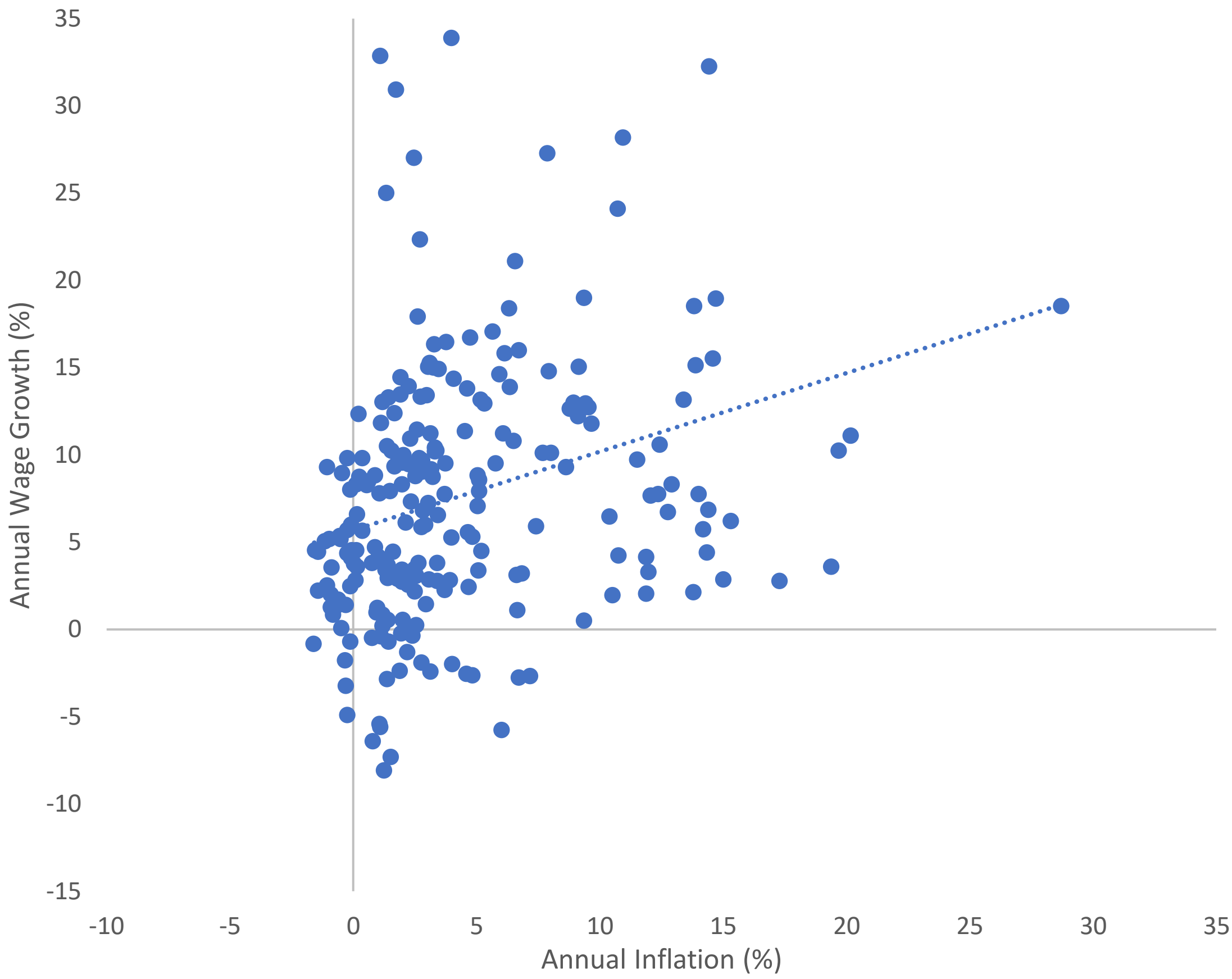

*Source: Author's calculations.*

Figure 2 presents the distribution of nominal wage growth across country-year observations. Three features stand out. First, most observations are concentrated in the low positive range, indicating that moderate nominal wage increases are the dominant outcome. Second, the left tail is thinner than the right tail, suggesting that nominal wage reductions are relatively infrequent and generally limited in magnitude. Third, the distribution is asymmetric, with occasional large upward adjustments but comparatively few substantial nominal wage declines.

This pattern is consistent with aggregate asymmetry in nominal wage adjustment, although the evidence should be interpreted cautiously given that the data are at the aggregate rather than micro level. Even so, the histogram suggests that nominal wages adjust unevenly across episodes and are rarely driven by large downward corrections. This does not by itself establish that rigidity amplifies inflation persistence. Rather, it indicates that internal nominal adjustment is constrained and non-symmetric, leaving open the question of whether such constraints ultimately strengthen or weaken inflation propagation in this setting.

**Figure 2. Histogram of Nominal Wage Changes**

Number of observations

60
50
40
30
20
10
0

≤ -19.0
(-19.0, -16.0]
(-16.0, -13.0]
(-13.0, -10.0]
(-10.0, -7.0]
(-7.0, -4.0]
(-4.0, -1.0]
(-1.0, 2.0]
(2.0, 5.0]
(5.0, 8.0]
(8.0, 11.0]
(11.0, 14.0]
(14.0, 17.0]
(17.0, 20.0]
(20.0, 23.0]
(23.0, 26.0]
(26.0, 29.0]
(29.0, 32.0]
(32.0, 34.0]
> 34.0

Wage Growth Brackets (%)

*Source: Author's calculations.*

Figure 3 documents substantial heterogeneity in realized exchange-rate volatility across transition economies, calculated through the standard deviation of exchange rate movements over the observing period. While some countries exhibit very low volatility, consistent with tightly managed or de facto fixed arrangements, others display much greater nominal flexibility. Transition economies therefore do not operate under a common external adjustment regime; they span a spectrum from externally constrained to externally flexible environments.

Importantly, exchange-rate volatility does not mechanically coincide with wage volatility. Some economies combine limited exchange-rate volatility with moderate wage volatility, others display the reverse configuration, and a subset exhibits low volatility on both margins. This suggests that internal and external nominal flexibility are distinct dimensions of adjustment rather than different expressions of the same institutional condition. For the purposes of the analysis, this distinction is important because the inflationary consequences of shocks may depend not only on whether adjustment is constrained, but also on which margin is constrained.

Figure 4. Wage and Exchange Rate Volatility (2011-2024)

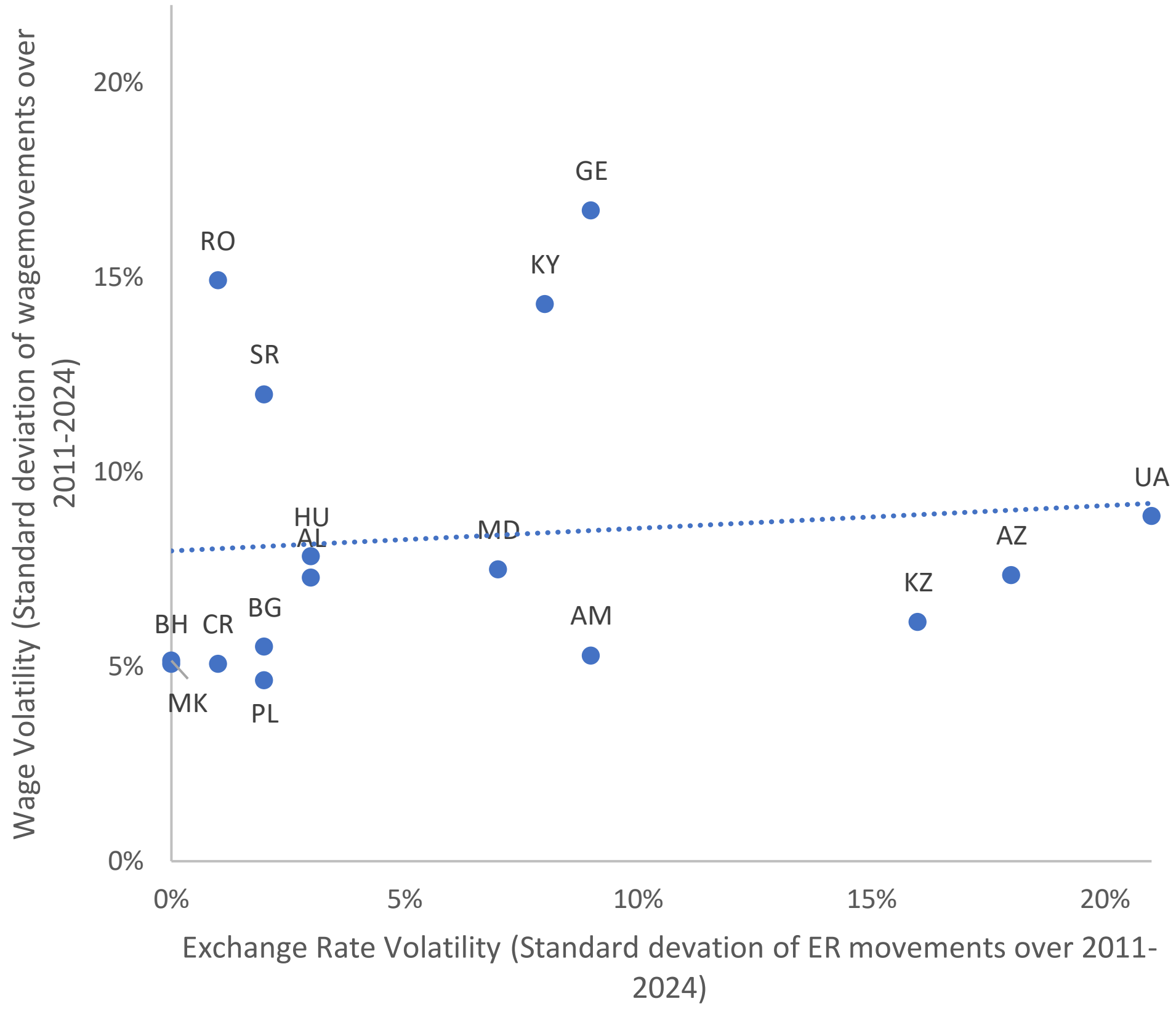

*Source: Author's calculations.*

These descriptive patterns suggest that transition economies differ substantially in the degree and composition of nominal adjustment. Wage responses to inflation are potentially incomplete and heterogeneous, aggregate wage adjustment is asymmetric, and exchange-rate behaviour varies markedly across countries. These stylised facts do not establish the sign of the relationship between nominal rigidity and inflation persistence ex ante. They do, however, indicate that the benchmark mechanism of uniform wage–price transmission is unlikely to apply straightforwardly in this setting. This motivates the formal analysis that follows, which examines whether internal and external rigidities amplify inflation persistence in the conventional way or instead condition it differently in transition economies.

## 4. Methodology and Data

### 4.1. Institutional measurement

To identify the wage rigidity, the analysis requires an institutional measure of how strongly wage formation is constrained from responding to shocks. In the benchmark case, nominal wage adjustment is informative about persistence only if wage-setting transmits past inflation into current wage claims through a sufficiently market-responsive mechanism. That assumption is unlikely to hold where wage formation is mediated by legal constraints that compress the scope for downward adjustment. The relevant institutional object is therefore not realized wage behaviour, but the legal structure governing wage-setting.

The Wage Rigidity Index (WRI) is constructed to capture that structure. It measures the extent to which wage outcomes are constrained by three institutional channels: the binding reach of collective bargaining outcomes beyond signatory parties, the degree of wage-setting centralization and hierarchical constraint across bargaining levels, and the rigidity of the statutory minimum-wage floor. The first channel captures the extent to which collectively negotiated wage standards apply only to signatories or also to non-members and non-signatory employers. Broader binding reach reduces firm-level discretion in wage-setting by extending negotiated pay conditions beyond the immediate parties to the agreement. The second channel captures whether higher-level bargaining outcomes restrict lower-level settlements and prevent lower-level bargaining from undercutting higher-level wage standards. The third captures the extent to which statutory minimum-wage rules impose a binding floor through coverage, rule-based updating, and non-regression provisions. Higher values of the index therefore indicate tighter institutional constraints on downward nominal wage adjustment.

The WRI is derived from AI-assisted coding of labour codes, collective-bargaining legislation, minimum-wage statutes, and related implementing provisions (Table 1). Legal texts are decomposed into standardized conceptual dimensions based on semantic content and legal enforceability, and then mapped into ordered indicators through a transparent coding protocol. Because these legal provisions are typically slow-moving, the resulting measure is treated as time-invariant over the sample window. Identification therefore relies on cross-country heterogeneity in the institutional structure of wage-setting rather than on within-country legal change.

**Table 1. Conceptual structure of Wage Rigidity Index (WRI)**

| Dimension ID | Dimension name | Range | Weight |
|---|---|---|---|
| **WRI_D1** | CBC & extension / binding reach | 0–2 | 33.3% |
| **WRI_D2** | Wage-setting centralisation / hierarchy | 0–2 | 33.3% |
| **WRI_D3** | Minimum wage rigidity | 0–2 | 33.3% |
| **WRI_TOTAL** | Wage Rigidity Index | 0–6 | 100% |

*Source: Authors' work.*

A broader institutional qualification is introduced through the Labour Protection Rigidity Index (LPRI) (**Table 2**). The baseline mechanism concerns wage formation narrowly defined, so the WRI is the appropriate primary measure. But if the estimated relationship reflects a wider legal architecture of labour-market regulation rather than the wage-setting channel specifically, it should also appear under a broader institutional proxy. The LPRI serves this purpose. It captures four dimensions of collective labour regulation: anti-union discrimination protection, trade-union rights and facilities, protection against employer or state interference in union activity, and dispute-resolution procedural rigidity. The LPRI is therefore not a wage-rigidity measure in the strict sense. It captures the broader institutional density of labour protection. Substituting the LPRI for the WRI allows the analysis to distinguish between a channel operating through wage formation specifically and one operating through the wider legal structure of collective labour relations.

**Table 2. Conceptual structure of Labor Protection Rigidity Index (LPRI)**

| Dimension ID | Dimension name | Range | Weight |
|---|---|---|---|
| **LPRI_D1** | Anti-union discrimination protection | 0–2.5 | 25% |
| **LPRI_D2** | Trade union rights & facilities | 0–2.5 | 25% |
| **LPRI_D3** | Protection against interference | 0–2.5 | 25% |
| **LPRI_D4** | Dispute-resolution procedural rigidity | 0–2.5 | 25% |
| **LPRI_TOTAL** | Labour Protection Index | 0–10 | 100% |

*Source: Authors' work*

The external adjustment margin is captured by exchange-rate regime rigidity. Under the conventional view, more rigid regimes limit nominal exchange-rate adjustment and shift shock absorption onto domestic prices. That implication depends, however, on treating regime choice purely as a relative-price constraint. Where exchange-rate arrangements also perform a credibility function, the relevant object is the effective rigidity of the regime as implemented rather than its official classification. For that reason, exchange-rate regime rigidity is measured using the de facto consensus classification of Couharde and Grekou (2024), which reconciles multiple behaviour-based regime taxonomies. Following the broader logic of Levy-Yeyati and Sturzenegger, the measure is inferred from observed policy behaviour rather than official declarations, while reducing classification noise and spurious regime switching relative to earlier de facto approaches. Unlike the WRI, the exchange-rate measure is allowed to vary discretely over time when regime shifts are identified in the underlying classification.

### 4.2. Empirical specification

Inflation dynamics are modeled using a dynamic panel framework that allows the degree of inflation persistence to vary systematically with institutional characteristics. The baseline specification is given by:

$$\pi_{i,t} = \alpha_i + \lambda_t + \rho_0 \pi_{i,t-1} + \rho_1(\pi_{i,t-1} \times WRI_i) + \rho_2(\pi_{i,t-1} \times ER_i) + \boldsymbol{X}'_{i,t}\ \beta + \varepsilon_{i,t} \qquad (1)$$

where $\pi_{i,t}$ denotes inflation in country i at time t; $\alpha_i$ and $\lambda_t$ capture country and time fixed effects; $WRI_i$ is the wage rigidity index; $ER_i$ is a measure of exchange rate regime rigidity. The vector $\boldsymbol{X}'_{i,t}$ includes standard macroeconomic controls, namely GDP growth, external price pressures proxied by changes in the terms of trade, and energy price shocks.

The coefficient $\rho_0$ captures baseline inflation persistence, while $\rho_1$ and $\rho_2$ measure how this persistence varies with wage rigidity and exchange rate regime rigidity, respectively. In this framework, institutional characteristics do not directly affect the level of inflation; rather, they shape the **propagation mechanism**, i.e. the extent to which past inflation feeds into current inflation.

Two testable hypotheses follow from this specification:

1. $\rho_1>0$: inflation persistence is higher in economies with higher wage rigidity, reflecting slower wage and cost adjustment.
2. $\rho_2>0$: inflation persistence is higher in economies with more rigid exchange rate regimes, reflecting reduced capacity for nominal adjustment through the exchange rate.

However, these predictions are not unambiguous in the transition-economy context. If wage rigidity reflects institutionalized or rule-bound wage-setting that weakens backward-looking indexation, then $\rho_1 \neq 0$ may arise. Likewise, if more rigid exchange-rate regimes operate as nominal anchors that discipline expectations, then $\rho_2 \neq 0$ may also obtain. The empirical analysis therefore evaluates whether institutional rigidities amplify inflation persistence, as in the standard benchmark, or instead dampen the propagation of inflationary shocks in transition economies.

Under more flexible institutional settings—either in labor markets or exchange rate regimes—the transmission of shocks is expected to be more short-lived, resulting in lower inflation persistence.

### 4.3. Identification and interpretation

Identification of (1) relies on the interaction between **within-country time variation in inflation** and **cross-country heterogeneity in institutional structures**. Because labor market institutions evolve slowly by design and exchange rate regimes tend to change discretely rather than continuously, the empirical strategy adopts a **cohort-based interpretation** of inflation dynamics.

Under this interpretation, countries can be viewed as belonging to implicit cohorts defined by their institutional configurations—specifically, the degree of wage rigidity and exchange rate regime rigidity. Rather than relying on frequent within-country regime changes, identification is achieved by comparing how inflation persistence differs across these cohorts when exposed to broadly similar macroeconomic conditions.

This approach shifts the focus from the incidence of shocks to their **propagation mechanisms**. While countries may experience comparable external shocks—such as energy price increases or changes in global trade conditions—the speed and persistence of inflation adjustment depend on the institutional environment through which these shocks are transmitted. In particular, rigid labor markets may delay wage and price adjustment, while rigid exchange rate regimes limit the role of nominal exchange rate movements in absorbing external imbalances.

To address the dynamic nature of the model and potential endogeneity of regressors, the specification is estimated using a generalized method of moments (GMM) estimator that exploits internal instruments derived from lagged variables. In particular, lagged inflation is treated as endogenous, given the well-known bias arising from its correlation with the fixed effects in dynamic panel settings.

In addition, selected regressors are treated as endogenous or predetermined to account for potential simultaneity and reverse causality. Real economic activity, proxied by GDP growth, may respond contemporaneously to inflation shocks through demand, monetary policy, and stabilization channels, justifying its treatment as endogenous. More importantly, the interaction terms between lagged inflation and institutional variables—capturing wage rigidity and exchange rate regime rigidity—are also treated as endogenous. While the institutional indices themselves evolve slowly and are largely predetermined, their interaction with lagged inflation reflects equilibrium outcomes that may be jointly determined with inflation dynamics.

In particular, exchange rate regimes are not exogenously assigned but are often adopted as policy responses to inflationary environments. A well-established strand of the literature emphasizes that countries facing high or unstable inflation may choose more rigid exchange rate arrangements—such as pegs or currency anchors—as a commitment device to stabilize prices and anchor expectations (e.g., Calvo and Reinhart, 2002; Fischer, 2001). This implies that exchange rate regime rigidity may be endogenous to inflation dynamics. Similarly, wage-setting institutions may adjust in response to persistent inflation through mechanisms such as indexation, backward-looking expectations, or collective bargaining responses (e.g., Blanchard and Katz, 1999; Galí, 2015). As a result, the interaction terms between lagged inflation and these institutional features may be correlated with the error term, warranting instrumentation.

By contrast, external price variables—namely energy price changes and terms-of-trade shocks—are treated as exogenous, as they are largely driven by global market conditions and are plausibly independent of country-specific inflation innovations.

The GMM estimator therefore relies on lagged levels of the endogenous variables as instruments for their differenced counterparts, with the instrument set restricted through lag limits and collapsing to mitigate instrument proliferation. This approach ensures consistent estimation while preserving a parsimonious and well-conditioned instrument matrix.

Overall, the empirical framework is consistent with theoretical models in which inflation persistence emerges endogenously from the interaction between nominal rigidities and policy constraints, rather than being imposed as a fixed structural parameter. By embedding institutional characteristics directly into the persistence mechanism, the analysis provides a structured way to assess how different policy and institutional environments shape the dynamics of inflation.

### 4.4. Capturing uncertainty

In contrast to standard macroeconomic variables, which are directly observed and quantitatively measured, the institutional indicators used in this analysis are subject to non-trivial measurement uncertainty. The wage rigidity and labor protection rigidity indices are derived from legal texts using an AI-assisted classification procedure. While this approach enables a detailed and comparable mapping of institutional features across countries, it inevitably introduces imprecision in translating qualitative legal provisions into quantitative indicators. Such imprecision may arise from differences in legal interpretation, enforcement, and the economic relevance of specific provisions. In this setting, classical measurement error is likely to attenuate estimated coefficients and weaken statistical significance, implying that the weak, unstable, and occasionally sign-reversing estimates observed for the wage rigidity interaction term in Table 3 may partly reflect attenuation bias rather than true parameter instability.

A related, though conceptually distinct, source of uncertainty applies to the exchange rate regime indicator. The classification used in this paper follows the synthesis approach of Couharde and Grekou (2024), which improves upon earlier de facto classifications by reconciling multiple methodologies and reducing classification noise. However, despite these advances, classification uncertainty cannot be fully eliminated. By construction, de facto regime identification relies on observable policy outcomes—such as exchange rate volatility and reserve movements—that are inherently noisy and may reflect multiple underlying policy behaviors. As a result, regime assignments should be interpreted as estimates of an underlying latent policy stance, rather than as perfectly observed variables.

To address these concerns, the analysis adopts a simulation-based approach that explicitly incorporates uncertainty in both institutional variables and propagates it through the estimation procedure. For wage rigidity, the observed index is treated as a point estimate of an underlying latent parameter, and alternative realizations are generated using the uncertainty bounds captured by the auxiliary band variables. For each draw, the interaction term between lagged inflation and wage rigidity is reconstructed, and the full dynamic panel model is re-estimated using the same specification and instrument set. 500 repetitions are used.

For exchange rate regimes, uncertainty is treated as probabilistic misclassification. Specifically, for each simulation, regimes are randomly reassigned according to the probabilities provided by the underlying classification, and the corresponding rigidity measure is recomputed. This ensures that uncertainty in regime identification is incorporated in a manner consistent with the categorical nature of the variable.

Together, this framework yields empirical distributions of the coefficients associated with the institutional interaction terms, allowing us to assess the sensitivity of the estimated persistence effects to plausible variation in the underlying institutional measures. Importantly, the approach treats measurement uncertainty as a feature of the data-generating process rather than as a weighting problem, thereby preserving the original econometric specification while allowing inference to account for imprecision in institutional variables.

### 4.5. Data

The empirical analysis covers a maximum of 20 transition economies from Central, Eastern and Southeastern Europe and the Commonwealth of Independent States over the period 2013–2024. The sample is defined based on the availability of legal information to create the wage and labor protection rigidity indices, which implies that four and one transition economies, respectively, drop out of this analysis and reduce the sample to 16 and 19 countries where both wage rigidity indices are used. The time window captures multiple inflationary episodes, including the post-pandemic inflation surge, while maintaining relative stability in labor market institutions. The sample thus provides sufficient within-country variation in inflation dynamics, alongside meaningful cross-country heterogeneity in institutional configurations.

The key dataset underlying this study refers to wage and labor protection rigidity indices. These are created by the authors, as per the description in Section 4.1. The exchange rate regime classification is taken from Couharde and Grekou (2024). The rest of the data to estimate (1) are taken from the World Development Indicators. Further descriptions of the variables are provided in Tables 3 and 4.

**Table 3 – Variables' definitions and sources**

| Variable | Definition | Source |
|---|---|---|
| **Inflation** | Annual growth of the Consumer Price Index in %. | World Development Indicators. |
| **Lagged inflation** | Lagged inflation. | World Development Indicators. |
| **Wage rigidity index** | Higher values imply higher rigidity. | Own calculations. |
| **Wage rigidity index band** | Band around the point estimate capturing the uncertainty / measurement error in decoding legal text into numerical value. | Own calculations. |
| **Labor protection rigidity index** | Higher values imply higher rigidity. | Own calculations. |
| **Labor protection rigidity index band** | Band around the point estimate capturing the uncertainty / measurement error in decoding legal text into numerical value. | Own calculations. |
| **Exchange rate regime** | Inverse values taken from 1 = Fix to 3 = Free float. Higher values imply higher rigidity. | Couharde and Grekou (2024). |
| **Import prices** | Inverse of the Net barter terms of trade index. Higher values imply higher import prices than export prices, i.e. worse terms of trade. | World Development Indicators. |
| **Energy prices** | Energy price index. | World Bank Commodity Price Data (The Pink Sheet). |
| **GDP growth** | Annual growth of GDP in %. | World Development Indicators. |

**Table 4 – Descriptive characteristics**

| Variable | Obs | Mean | Std.Dev. | Min | Max |
|---|---|---|---|---|---|
| **Inflation** | 251 | 4.62 | 5.59 | -1.5 | 48.7 |
| **Wage rigidity index** | 213 | 2.93 | 1.34 | 1.0 | 5.5 |
| **Labor protection rigidity index** | 253 | 6.75 | 0.72 | 5.8 | 8.0 |
| **Exchange rate regime** | 252 | 6.48 | 0.78 | 1 | 3 |
| **Import prices (% change)** | 220 | 0.00 | 0.09 | -0.5 | 0.4 |
| **Energy prices (% change)** | 231 | -0.02 | 0.35 | -0.6 | 0.6 |
| **GDP growth** | 252 | 2.82 | 4.28 | -28.8 | 13.9 |

## 5. Results and discussion

### 5.1. Baseline results

Table 5 presents the baseline estimates of equation (1), examining how inflation persistence varies with institutional characteristics governing internal and external adjustment. The table reports both fixed-effects and dynamic panel specifications across alternative samples, allowing for a comprehensive assessment of the relationship between institutional rigidities and inflation dynamics. At the bottom of the table, the diagnostic tests are broadly supportive of the empirical strategy: Hansen test p-values indicate that the instrument sets are not rejected, while some sensitivity in the AR(2) test suggests that the dynamic specifications should be interpreted with appropriate caution.

Across specifications, inflation exhibits a clear degree of persistence, as indicated by the positive and statistically significant coefficient on lagged inflation. This confirms that past inflation contributes to current inflation, consistent with gradual nominal adjustment. The persistence coefficient is generally larger in the dynamic specifications, which is expected, as these estimators account for endogeneity and attenuation bias in lagged inflation, leading to less downward-biased estimates relative to static models.

Turning to the key variable of interest, the interaction between lagged inflation and the wage rigidity index reveals a pattern that is both informative and non-trivial. In the fixed-effects specifications, the coefficient is weakly positive, which is broadly in line with the conventional view that rigid wage-setting slows cost adjustment and thereby sustains inflation persistence. Once dynamics and endogeneity are accounted for, however, the sign turns negative and statistically significant. This suggests that, in the dynamic setting, higher wage rigidity is associated with lower inflation persistence rather than higher persistence. One possible interpretation is that more rigid or institutionalized wage-setting frameworks limit the transmission of inflation shocks into wages, thereby weakening second-round effects and dampening the wage–price feedback mechanism. In this sense, wage rigidity may act not only as a friction but also as a stabilizing device, reducing the propagation of inflationary pressures. This reading is consistent with the non-monotone role of real wage rigidities emphasized by Blanchard and Galí (2007), according to which wage rigidities may, under certain configurations, dampen the response of marginal costs and thereby weaken inflation propagation.

A similar pattern emerges for exchange-rate regime rigidity. In the fixed-effects specifications, the interaction between lagged inflation and exchange-rate rigidity is positive, and in one case statistically significant, which is consistent with the benchmark argument that limited exchange-rate flexibility constrains external adjustment and shifts the burden of shock absorption onto domestic prices. In the dynamic specifications, however, the sign becomes negative, and in the final specification the coefficient is statistically significant. This indicates that, once endogeneity is addressed, greater exchange-rate regime rigidity may also be associated with lower inflation persistence. This is in line with the view that rigid exchange-rate arrangements may operate as nominal anchors in low-credibility environments, disciplining expectations rather than merely constraining adjustment (Calvo and Reinhart, 2002; Calvo, Celasun and Kumhof, 2002). At the same time, this effect should still be interpreted with caution, as it is not uniformly strong across all specifications. The final specification points to a more systematic role of exchange-rate regime rigidity in conditioning inflation dynamics. More generally, the fact that both institutional interaction terms become negative in the dynamic setting suggests that institutional rigidities need not operate solely as frictions that delay adjustment; under certain configurations, they may also act as stabilizing constraints that weaken the propagation of inflationary shocks.

By contrast, the macroeconomic controls exhibit stable and economically meaningful patterns. Changes in import prices and energy prices are positively associated with inflation and are statistically significant in most specifications, highlighting the central role of external price shocks in shaping inflation dynamics in transition economies, which are typically small and highly open. GDP growth enters with a negative coefficient, suggesting that stronger economic performance may coincide with lower inflation, potentially reflecting supply-side improvements or stabilization episodes.

Table 5 – Baseline results

| | Dependent variable: Inflation rate | | | | | | | |
|---|---|---|---|---|---|---|---|---|
| | Fixed effects | | | | Difference-GMM | | | |
| | (1) | (2) | (3) | (4) | (5) | (6) | (7) | (8) |
| **Lagged inflation** | 0.391*** | 0.406*** | 0.246** | 0.303** | 0.591*** | 1.699*** | 0.785** | 3.729*** |
| | (0.075) | (0.082) | (0.087) | (0.118) | (0.043) | (0.659) | (0.331) | (1.271) |
| **Lagged inflation * WRI** | | 0.0951 | | 0.115** | | -0.409** | | -0.567* |
| | | (0.069) | | (0.053) | | (0.172) | | (0.332) |
| **Lagged inflation * exchange rate regime** | | | 0.131** | 0.0525 | | | -0.106 | -0.663** |
| | | | (0.058) | (0.056) | | | (0.136) | (0.336) |
| **Change in import prices** | 6.367** | 5.264** | 6.064* | 5.226* | 5.246* | 4.829 | 5.950* | 4.039 |
| | (2.978) | (2.433) | (2.927) | (2.508) | (3.159) | (3.991) | (3.521) | (3.633) |
| **Change in energy prices** | 6.370*** | 7.412*** | 6.394*** | 7.473*** | 6.969*** | 12.18*** | 7.017*** | 12.17*** |
| | (1.049) | (0.972) | (1.063) | (1.007) | (1.863) | (1.834) | (2.001) | (3.097) |
| **GDP growth** | -0.402*** | -0.394** | -0.385*** | -0.405** | -0.631*** | -1.240*** | -0.622*** | -1.345*** |
| | (0.071) | (0.134) | (0.070) | (0.141) | (0.202) | (0.241) | (0.228) | (0.426) |
| **Constant** | 4.246*** | 3.420*** | 4.063*** | 3.433*** | 3.992*** | 5.236*** | 4.209*** | 5.606*** |
| | (0.494) | (0.616) | (0.297) | (0.584) | (0.581) | (0.729) | (0.501) | (1.879) |
| | | | | | | | | |
| **Observations** | 219 | 175 | 219 | 175 | 219 | 175 | 219 | 175 |
| **R-squared** | 0.287 | 0.435 | 0.314 | 0.438 | | | | |
| **Number of idcountry** | 20 | 16 | 20 | 16 | 20 | 16 | 20 | 16 |
| **Arellano-Bond test for AR(2)** | | | | | 0.891 | 0.105 | 0.847 | 0.051 |
| **Hansen test for overid. restrictions (p-value)** | | | | | 0.085 | 0.246 | 0.0705 | 0.437 |

*Source: Authors' calculations. *, ** and *** refer to statistical significance at the 10%, 5% and 1% level, respectively. Standard errors provided in parentheses. Standard errors robust to arbitrary heteroskedasticity. Country and time fixed effects used.*

Overall, the evidence in Table 5 suggests that institutional rigidities play a limited and nuanced role in shaping inflation persistence. Both wage rigidity and exchange rate regime rigidity tend to exhibit **dampening effects on inflation persistence**, pointing to a potential role in weakening the propagation of inflationary shocks. However, these effects are not uniformly robust across models and should be interpreted with caution. By contrast, external price shocks—captured by import and energy prices—emerge as the dominant and most consistent drivers of inflation persistence in transition economies, underscoring the central importance of global factors in shaping domestic inflation dynamics in small and open economies.

### 5.2. Accounting for measurement uncertainty

The baseline results reported in Table 5 suggest a potential dampening role of nominal rigidities for inflation persistence. However, this finding must be interpreted with caution, as it may partly reflect measurement uncertainty in the construction of the institutional variables, rather than the absence or firm presence of an underlying economic relationship.

The simulation results reported in Table 6 provide a more disciplined assessment of the role of wage rigidity once measurement uncertainty is explicitly taken into account. While the baseline estimates indicate a sizeable negative coefficient (−0.567), the distribution of simulated coefficients is centered at lower magnitudes across all specifications, with mean values ranging between −0.293 and −0.068 and medians between −0.303 and −0.082. This confirms that the baseline magnitude is partly inflated by measurement noise.

Importantly, however, the **direction and statistical relevance of the effect remain largely preserved**. Across all specifications, the distribution of coefficients is clearly tilted toward negative values, as reflected in the low share of positive coefficients (particularly under the uniform distribution of the measurement error, which we prefer). More notably, a **large majority of simulated coefficients remain negative and statistically significant at the 5% level**, ranging from 70% to nearly 90% of the draws. This provides strong evidence that the dampening effect of wage rigidity on inflation persistence is not an artefact of measurement error, but rather a robust feature of the data.

At the same time, the dispersion of the estimates remains non-negligible, with standard deviations between 0.155 and 0.259, indicating that while the **sign and significance of the effect are stable, its magnitude is subject to uncertainty**. The results are also broadly consistent across alternative assumptions regarding the distribution of measurement error and when controlling for exchange rate regime rigidity, suggesting that the findings are not driven by distributional assumptions or multicollinearity between institutional variables.

**Table 6 – Simulated results for wage rigidity index**

| | | Uniform distribution | Normal distribution | ERR variable not excluded |
|---|---|---|---|---|
| **Baseline** | Coefficient | -0.567 | -0.567 | -0.409 |
| | Standard error | 0.332 | 0.332 | 0.172 |
| | AR(2) p-value | 0.051 | 0.051 | 0.021 |
| | Hansen p-value | 0.437 | 0.437 | 0.246 |
| **Simulated** | Mean of coefficients | -0.293 | -0.126 | -0.068 |
| | Median of the coefficients | -0.303 | -0.146 | -0.082 |
| | SD of the coefficients | 0.259 | 0.226 | 0.155 |
| | Share of positive coefficients | 0.114 | 0.256 | 0.296 |
| | Share of negative coefficients which are significant at 5% | 0.886 | 0.744 | 0.704 |
| | AR(2) p-value | 0.107 | 0.144 | 0.104 |
| | Hansen p-value | 0.213 | 0.174 | 0.147 |

*Source: Authors' calculations.*

Turning to exchange rate regime rigidity, Table 7 reports the results of an analogous simulation exercise that accounts for classification uncertainty in the de facto regime assignment. The results now point to a **strong and highly robust pattern**. In the specification including both wage rigidity and exchange rate rigidity, the baseline coefficient is negative (−0.663), and the simulated distribution is not only preserved but shifts toward **larger negative values**, with a mean of −1.668 and a median of −1.747. The distribution is almost entirely concentrated on the negative side, with virtually no positive realizations (0.4%). Moreover, an overwhelming share of simulated coefficients—close to 100%—remain **negative and statistically significant at the 5% level**, providing compelling evidence of a robust dampening effect of exchange rate regime rigidity on inflation persistence.

In contrast, when wage rigidity is excluded from the specification, the estimated effect of exchange rate rigidity becomes much smaller in magnitude, with a baseline coefficient of −0.106 and a simulated mean of −0.130. The dispersion is also substantially lower (standard deviation of 0.056), and the share of positive coefficients is effectively zero. At the same time, the coefficient remains consistently negative and statistically significant across all simulations. This comparison suggests that while the presence of wage rigidity amplifies the magnitude of the estimated effect, the dampening role of exchange rate rigidity is **independently present and highly stable**, even in more parsimonious specifications.

**Table 7 – Simulated results for exchange rate regime**

| | | Specification - WRI and ERR | Specificatio n – WRI not included |
|---|---|---|---|
| **Baseline** | Coefficient | -0.663 | -0.106 |
| | Standard error | 0.336 | 0.136 |
| | AR(2) p-value | 0.051 | 0.847 |
| | Hansen p-value | 0.437 | 0.075 |
| **Simulated** | Mean of coefficients | -1.668 | -0.130 |
| | Median of the coefficients | -1.747 | -0.125 |
| | SD of the coefficients | 0.656 | 0.056 |
| | Share of positive coefficients | 0.004 | 0.000 |
| | Share of negative coefficients which are significant at 5% | 0.996 | 1.000 |
| | AR(2) p-value | 0.268 | 0.843 |
| | Hansen p-value | 0.137 | 0.095 |

***Source: Authors' calculations.***

The results from Tables 6 and 7 indicate that both wage rigidity and exchange rate regime rigidity contribute to dampening inflation persistence, but with different degrees of robustness. While the effect of wage rigidity is present but sensitive to measurement assumptions, the effect of exchange rate regime rigidity emerges as **systematically negative, statistically robust, and remarkably stable across simulations**.

Importantly, incorporating measurement and classification uncertainty does not weaken this conclusion. In the case of exchange rate regimes, the simulation exercise reinforces the baseline finding by demonstrating that the negative effect persists—and even strengthens—across alternative admissible realizations of the underlying classification. This suggests that the observed relationship is not an artefact of measurement noise, but rather reflects a genuine structural feature of inflation dynamics.

These results should nonetheless be interpreted with some caution. Although the direction and statistical significance of the effect are highly stable—particularly for exchange rate regimes—the dispersion of the simulated coefficients in the full specification remains non-negligible, indicating uncertainty about the precise magnitude. In addition, some sensitivity in specification diagnostics persists. Accordingly, the findings support an **existing but variable dampening role of wage rigidity** and a **robust dampening role of exchange rate rigidity**, while suggesting that its quantitative impact may vary across institutional and macroeconomic contexts.

### 5.3. Robustness to alternative measurements

As a robustness check, we replace the wage rigidity index (WRI) with a broader labor protection rigidity index (LPRI), which captures a wider set of institutional features governing employment relationships, including dismissal protection, contractual constraints, and collective bargaining arrangements. While not a direct measure of wage-setting rigidity, LPRI can be interpreted as a proxy insofar as stronger employment protection is typically associated with more rigid wage-setting environments and reduced flexibility in labor cost adjustment.

The results reported in Table 8 broadly confirm the main findings. In the dynamic specifications, the interaction between lagged inflation and labor protection rigidity remains **negative and statistically significant**, with coefficients of −0.0176 and −0.104, the latter significant at 10%. This pattern is consistent with the baseline results and reinforces the interpretation that institutional rigidities in the labor market tend to **dampen inflation persistence**, rather than amplify it.

At the same time, the magnitude of the estimated effects is smaller and less stable compared to the baseline specification. In particular, the coefficient in column (3) is negative but not statistically significant, indicating some sensitivity to the precise measurement of institutional rigidity. This is expected, given that LPRI captures broader labor market features that extend beyond wage-setting mechanisms per se, and therefore constitutes a noisier proxy for the channel of interest.

The results for exchange rate regime rigidity remain qualitatively consistent. Although the coefficient is positive in the fixed-effects specification, it turns negative in the dynamic specification, in line with the baseline findings, albeit without statistical significance. This suggests that the dampening role of external rigidity is not driven by the specific choice of the wage rigidity measure, even if its precision varies across specifications.

Overall, the robustness exercise supports the central conclusion of the paper: institutional rigidities—both in labor markets and exchange rate arrangements—tend to **weaken the propagation of inflation shocks**, although the strength and precision of this effect depend on how these rigidities are measured.

**Table 8 – Robustness results for wage rigidity**

| | Fixed effects | | Difference GMM | |
|---|---|---|---|---|
| | (1) | (2) | (3) | (4) |
| **Lagged inflation** | -0.185 | 0.169 | 1.021 | 0.989** |
| | (0.539) | (0.461) | (0.718) | (0.406) |
| **Lagged inflation * LPRI** | 0.0862 | -0.0176* | -0.0705 | -0.104* |
| | (0.074) | (0.086) | (0.107) | (0.074) |
| **Lagged inflation * exchange rate regime** | | 0.180 | | 0.126 |
| | | (0.105) | | (0.078) |
| **Change in import prices** | 5.662* | 5.281* | 2.913 | 2.623 |
| | (2.879) | (2.786) | (3.826) | (2.231) |
| **Change in energy prices** | 6.331*** | 6.411*** | 7.101*** | 6.194*** |
| | (1.118) | (1.082) | (1.050) | (1.188) |
| **GDP growth** | -0.396*** | -0.387*** | -0.473*** | -0.390*** |
| | (0.070) | (0.073) | (0.067) | (0.097) |
| **Constant** | 4.328*** | 4.183*** | 3.849*** | 3.566*** |
| | (0.449) | (0.310) | (0.499) | (0.573) |
| | | | | |
| **Observations** | 208 | 208 | 208 | 208 |
| **R-squared** | 0.284 | 0.316 | | |
| **Number of idcountry** | 19 | 19 | 19 | 19 |
| **Arellano-Bond test for AR(2)** | | | 0.995 | 0.836 |
| **Hansen test for overid. restrictions (p-value)** | | | 0.773 | 0.142 |

*Source: Authors' calculations. *, ** and *** refer to statistical significance at the 10%, 5% and 1% level, respectively. Standard errors provided in parentheses. Standard errors robust to arbitrary heteroskedasticity. Country and time fixed effects used.*

## 6. Conclusion

This paper examined the role of institutional rigidities in shaping inflation persistence in transition economies, focusing on the interaction between internal adjustment mechanisms, governed by labor market institutions, and external adjustment mechanisms, governed by exchange rate regimes. The central objective was to move beyond treating inflation persistence as a fixed structural parameter and instead analyze it as an endogenous outcome conditioned by institutional constraints.

The analysis was conducted on a large panel transition economies over the period 2013–2024, combining newly constructed indices of wage rigidity and labor protection—derived from AI-assisted coding of legal texts—with de facto measures of exchange rate regime rigidity. These data were integrated with standard macroeconomic indicators on inflation, external prices, and economic activity.

Methodologically, the paper adopts a dynamic panel framework in which inflation persistence is allowed to vary with institutional characteristics through interaction terms. Estimation was performed using a GMM approach to address endogeneity and dynamic panel bias. Importantly, the analysis follows a **cohort-based interpretation of identification**, whereby countries are implicitly grouped according to their institutional configurations. Rather than relying on frequent within-country institutional changes, identification is achieved by comparing how inflation persistence differs across these institutional cohorts when exposed to similar macroeconomic conditions. Measurement and classification uncertainty in the institutional variables are explicitly incorporated through simulation-based techniques, allowing inference to account for imprecision in their construction.

The results yield three main findings. *First*, inflation persistence is a robust feature of transition economies, but its magnitude varies systematically across institutional configurations, consistent with the cohort-based interpretation. *Second*, both wage rigidity and exchange rate regime rigidity tend to **dampen inflation persistence**, suggesting that institutional constraints may weaken the transmission of past inflation into current price dynamics. This effect is particularly strong and robust for exchange rate regime rigidity, which consistently exhibits a negative and statistically significant relationship with inflation persistence across simulations, while the effect of wage rigidity, though present, is more sensitive to measurement assumptions. *Third*, accounting for measurement uncertainty reduces the magnitude of estimated effects but does not overturn their direction or significance, reinforcing the interpretation that these relationships reflect underlying structural mechanisms rather than artefacts of measurement noise.

These findings carry several implications. From a policy perspective, they suggest that institutional rigidities should not be viewed solely as sources of inefficiency or adjustment frictions, but also as potential stabilizing mechanisms that can dampen second-round inflationary effects. In particular, exchange rate arrangements may play a critical role as nominal anchors in environments characterized by weaker monetary credibility. More broadly, the results point to the importance of considering the joint configuration of internal and external adjustment margins when designing macroeconomic stabilization frameworks.

From a research perspective, the paper contributes by integrating institutional analysis into the study of inflation dynamics and by proposing a framework that explicitly incorporates measurement uncertainty in institutional variables. Future work could extend this approach by exploring nonlinearities, regime-dependent effects, and interactions with monetary policy frameworks, as well as by further refining the measurement of institutional rigidity.